\documentclass{elsart}

 \usepackage{epsfig}

\begin{document}

\begin{frontmatter}

 \title{Piezoelectric mechanism for the orientational pinning of bilayer Wigner crystals and stripes
 in a GaAs matrix }

 \author{D. V. Fil}
%\ead{fil@isc.kharkov.com}

\address{Institute for Single Crystals National Academy of Sciences, Lenin av. 60, 61001,
Kharkov, Ukraine}

\begin{abstract}
 We investigate the phonon mechanism for the
 orientational pinning of Wigner crystals and stripes in
 two-dimensional electron layers in GaAs matrices.
 We find the orientation of bilayer Wigner
 crystals on the (001), (111), (0$\bar{1}$1) and (311) interfaces versus the
 interlayer distance and determine the regions of parameters, where polydomain
 structures can emerge. For the stripe states in electron layers situated close
 to the (001) surface we show that the interference between the
 piezoelectric  and deformation potential interaction may be responsible for the
 preferable orientation of the stripes along the [110] direction. For
 the bilayer system on the (001) interfaces we predict the
 suppression of the resistance anisotropy.
 \end{abstract}

\end{frontmatter}
\section{Introduction}
Due to the number of technological advantages AlGaAs
heterostructures are the most widely used systems to construct
two-dimensional electron layers. Low level of disorder in this
systems make it possible to realize the states with nonuniform
spatial distribution of electrons. The formation of these states
is regulated by the Coulomb interaction, which, under certain
conditions, forces the transition into the Wigner crystal or the
stripe state. Since in cubic crystals the Coulomb interaction is
the isotropic one,  the energy of such states does not depend, in
the first approximation, on their orientation relative to the
crystallography axes of the host matrix. Nevertheless,
experimental studies of the resistance anisotropy in high Landau
levels \cite{1,2,11} (which is considered as an indication of the
stripe state) show that certain orientational symmetry-breaking
mechanisms exist.

In this paper we consider the electron-phonon interaction as a
possible origin of the native anisotropy. The main source of the
anisotropy is the piezoelectric interaction, which remains
anisotropic in cubic systems. This mechanism was studied by Rashba
and Sherman \cite{3,4}. In \cite{3,4} the influence of the
piezoelectricity on the structure and orientation of monolayer
Wigner crystals in isotropic piezoelectrics was investigated.
Since in GaAs crystals the anisotropy of the elastic moduli is
quite large, the results of \cite{3,4} cannot be applied directly
to them. In this paper, using the approach lightly different from
\cite{3,4}, we consider the model, where the anisotropy of the
elastic moduli of GaAs is taken into account. In section \ref{s2}
the model is applied to the study of orientation of bilayer Wigner
crystals. Technical details of the calculations were given in Ref.
\cite{5}. Here we mainly concentrate on the results. For the
bilayer systems on the (001), (111), (0$\bar{1}$1) and (311)
interfaces we find the stable and metastable orientations of the
Wigner crystal versus the interlayer separation and determine the
regions of parameters, where polydomain Wigner crystal structures
can be realized.

In section \ref{s3} we address the problem of stripe orientation
in high Landau levels. It was shown in \cite{6} that in an
electron layer parallel to the (001) crystallography plane the
piezoelectricity causes the preferential orientation of the
stripes along the [110] or the [$\bar{1}$10] direction, while the
piezoelectric mechanism alone does not explain further lowering of
the symmetry of the interaction (in magnetic field perpendicular
to the layer the [110] orientation is only observed). We extend
the model of Ref. \cite{6} and take into account  the
piezoelectric as well as the deformation potential interaction. It
is shown that for the electron layers situated near the surface of
the sample these two channel of the electron-phonon interaction
are strongly interfere and due to this effect the $C_{4v}$
symmetry of the electron-electron interaction is reduced to the
$C_{2v}$ one. We also analyze the orientation of the stripes in
the electron layers parallel to the (111), (0$\bar{1}$1) and (311)
crystallography planes, and in bilayer systems.

\section{Piezoelectric mechanism of orientation of bilayer Wigner crystals}
\label{s2}
 The  electrostatic potential $\varphi$ of an electron in a
piezoelectric matrix can be found from the solution of the
following equations:
\begin{equation}\label{1}
  \nabla {\bf D} = 4\pi e\delta({\bf r}),
\end{equation}
\begin{equation}\label{2}
  \frac {\partial \sigma_{ik}}{\partial x_k}=0,
  \end{equation}
  where
  $D_i=-\varepsilon\partial_i\varphi -
  4\pi\beta_{i,kl} u_{kl}$
is the electric displacement vector,
  $\sigma_{ik}=\lambda_{iklm}u_{lm}-\beta_{l.ik}
  \partial_l\varphi$,
the stress tensor, $\varepsilon$, the dielectric constant,
$\lambda_{iklm}$, the elastic moduli tensor, $\beta_{i,kl}$, the
piezoelectric moduli tensor, $u_{ik}$, the strain tensor.

Solving  Eqs. (\ref{1},\ref{2})  one can present the
electron-electron interaction potential  in the form:
\begin{equation}\label{3}
  V({\bf r}) =\frac{e^2}{\varepsilon r}-\chi\frac{e^2}{\varepsilon
  r}\sum_{n\geq 0}\sum_{|m|\leq n} A_{nm} Y_{nm}(\Theta_{\bf r},\phi_{\bf r})
  + O(\chi^2)
\end{equation}
where $Y_{nm}(\theta,\phi)$ are the spherical harmonics,
$\chi=e_{14}^2/\varepsilon c_{11}$ ( here and below the cubic
symmetry of the host matrix is assumed). For the case of an
anisotropic crystal ($c_{11}\ne c_{12}+2 c_{44}$) the coefficients
$A_{nm}$ are computed numerically.

The second term in r.h.s. of Eq. (\ref{3}) describes the second
order correction to the electron-electron interaction caused by
the virtual exchange of acoustic phonons. This  term contains the
dependence on the direction of the bond.

The lattice sum for the interaction (\ref{3}) can be reduced to
the rapidly convergent form (see Ref. \cite{5}). Using the rapidly
convergent form we have computed the energy of the bilayer Wigner
crystal and found its stable and metastable orientations  versus
the parameter $\eta=d\sqrt{n/2}$ ($d$ is the interlayer distance,
and $n$ is the electron density). In principle, the piezoelectric
interaction may influence on the symmetry of the electron lattice
as well, but  for GaAs, where the parameter $\chi$ is small
($\approx 2\cdot 10^{-4}$), this effect is negligible, and the
phase diagram is almost the same as in case of pure Coulomb
interaction \cite{8}.

\begin{figure}
\begin{center}
\centerline{\epsfig{figure=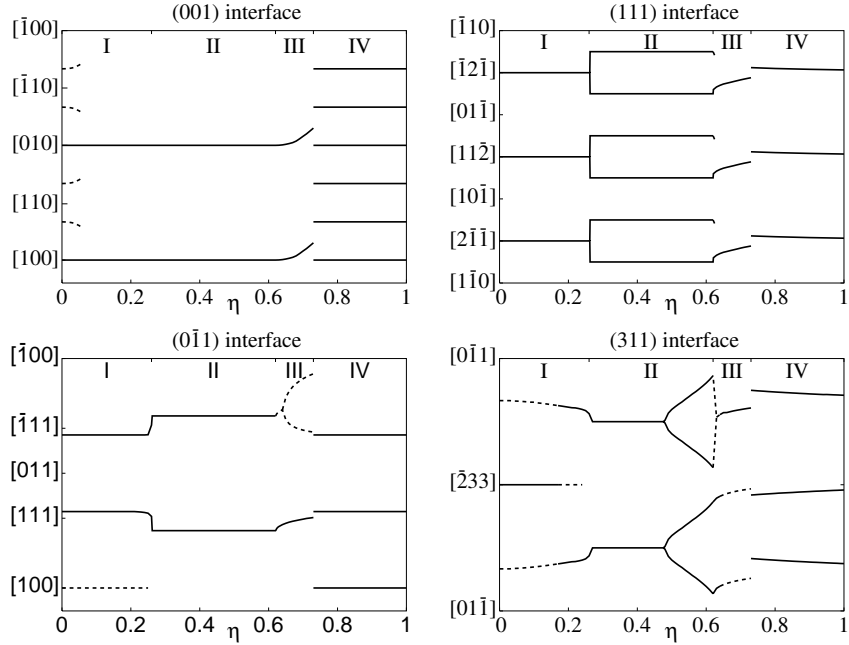,width=12cm}}
%\epsfbox{fdop1n.ps}
\end{center}
\caption{\label{f1}Direction of the shortest in-layer
electron-electron bond versus the interlayer separation for the
bilayer Wigner crystal on the (001), (111), (0$\bar{1}$1), and
(311) interfaces. Solid curves correspond to the stable
orientation; dotted curves, the metastable one. The Roman numbers
indicate the regions, where different types of the electron
lattice are realized: I, staggered rectangle; II, staggered cubic;
III - staggered rhombic; IV, staggered hexagonal (the
one-component hexagonal phase region is too narrow to be shown in
this scale) }
\end{figure}

For the calculations we use the elastic moduli of GaAs:
$c_{11}=12.3\cdot 10^{10}$ $N/m^2$, $c_{12}=5.7\cdot 10^{10}$
$N/m^2$, $c_{44}=6.0\cdot 10^{10}$ $N/m^2$. We would like to note
that the results do not change qualitatively under the 10\%
variation of $c_{ik}$. Therefore, one can neglect a small
difference of $c_{ik}$ in pure GaAs and in AlGaAs. The results of
the calculations for the (001), (111), (0$\bar{1}$1), and (311)
interfaces are presented in Fig. \ref{f1}, where the most
preferable orientation of the shortest primitive lattice vector is
shown. One can see that in general case the orientation of the
bilayer Wigner crystal is sensitive to the interlayer separation.
A common feature of the dependences presented is a jump-like
reorientation under the 1-st order type transition from the
rhombic to the hexagonal phase. For the  (111) and (0$\bar{1}$1)
interfaces a  step-like reorientation is also observed under the
2-nd order type transition from the rectangle to the cubic phase.

It is instructive to determine the regions, where a polydomain
structure of the bilayer Wigner crystal may emerge. As it follows
from Fig. \ref{f1} the polydomain structure is expected for the
rectangle, rhombic and hexagonal phases in the (001) oriented
bilayers, for the rectangle, square and rhombic phases in the
(111) bilayers, for the rectangle phase in the (0$\bar{1}$1)
bilayers, and for all phases (except the one-component hexagonal)
in the (311) bilayer. In the last case the rectangle  phase may
show such a behavior at $\eta>0.17$, and the cubic phase - at
$\eta>0.47$.

\section{Piezoelectric mechanism of stripe orientation}
\label{s3}
 In this section we consider the role of the electron-phonon
interaction in the orientation of stripe states in high Landau
levels. We consider an electron layer situated close to the (001)
surface of the sample and take into account two channel
(piezoelectric and deformation potential) of the electron-phonon
interaction. For the (001) surface the matrix elements of the
interaction of the electrons with surface acoustic modes are
in-phase for these two channels (see \cite{9}), and the phonon
mediated interaction contains the interference term. The
interference term may lower the symmetry of the electron-electron
interaction.

The potential energy is given by the expression
\begin{equation}
U=\int_{z<0} d^3 r ({{\bf E D}\over 8 \pi } +{u_{ik}
\sigma_{ik}\over 2 }) + \int_{z>0} d^3 r { E^2\over 8 \pi } +
U_{def} \ , \label{4}
\end{equation}
where {\bf E} is the electric field, and the deformation potential
interaction is chosen in the form
\begin{equation}
U_{def}=\int d^3 r \Lambda \rho (u_{xx}+u_{yy}) \delta(z+a) \ .
\label{5}
\end{equation}
Here $\Lambda$ is the deformation potential constant, $\rho$, the
electron density, $a$, the distance between the electron layer and
the surface. Since we consider the model of a zero-thickness
electron layer, the interaction with $u_{zz}$ deformations is not
included in (\ref{5}).

The quantities {\bf D} and $\sigma_{ik}$ satisfy the equations
$\nabla {\bf D}=0$, $\partial_k \sigma_{ik}=0$. The boundary
conditions at the free surface are the continuity of $D_z$ and
$E_x$, $E_y$, and the vanishing of $\sigma_{iz}$. At $z=-a$ the
quantities $D_z$ and $\sigma_{iz}$ are discontinuous:
\begin{equation}
D_z \Big|_{z=-a+0}-D_z\Big|_{z=-a-0} = 4 \pi e \rho \ , \label{6}
\end{equation}
\begin{equation}
\sigma_{iz}\Big|_{z=-a+0}-\sigma_{iz}\Big|_{z=-a-0}=-F_i \ ,
\label{7}
\end{equation}
where ${\bf F}= \Lambda (\partial_x \rho, \partial_y \rho,0)$ is
the tangential force applied to the unit area of the interface.
This force is induced by the deformation potential interaction. We
should note that discontinuity of the stresses is the consequence
of the zero-thickness approximation for the electron layer. Using
the boundary conditions specified one can reduce the energy
(\ref{4}) to the form
\begin{equation}
U={1\over 2}\int d^2 r (e \rho \varphi - u_i F_i) \ , \label{8}
\end{equation}
where  the electrostatic potential $\varphi$  and the displacement
field ${\bf u}$ are taken at $z=-a$. The energy (\ref{8}) can be
presented as a sum of a pure Coulomb part (which is isotropic) and
a phonon part $U_{\rm ph}$ (which includes the isotropic as well
as the anisotropic terms).

For simplicity, we consider the unimodal approximation for the
electron density modulation $\rho =\rho_0 \cos {\bf q r}_{pl}$,
where the absolute value of {\bf q} is determined by the period
$l$ of the stripe structure ($q=2\pi/l$) and its direction is
perpendicular to the axis along which the stripes are aligned.

\begin{figure}
\begin{center}
\centerline{\epsfig{figure=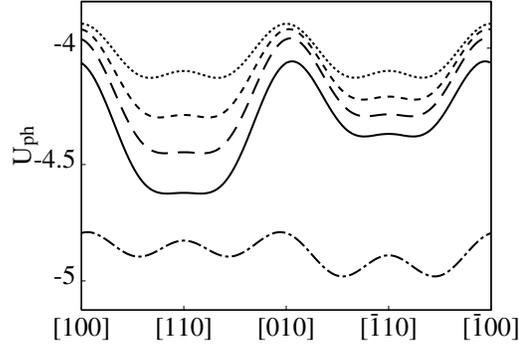,width=8cm}} %\epsfbox{s001.ps}
\end{center}
\caption{\label{f5} Phonon contribution to the energy of the
stripe phase (in units of $\chi U_c$) near the (001) surface
versus the direction of the stripes at $a/l=$1.0, 0.6, 0.5, 0.4,
0.2 (from top to bottom).}
\end{figure}

We concentrate on the anisotropic GaAs crystal and  compute the
values of $\varphi$  and $u_i$ at $z=-a$ numerically. We use the
parameters $e_{14}=0.15$ $C/m^2$, $\Lambda=7.4$ $eV$, $\varepsilon
=12.5$, $l=2\cdot 10^3$ \AA,  and the elastic moduli given above
for the calculation. For the (001) surface the dependence of the
energy $U_{\rm ph}$ on the direction of stripes is shown in Fig.
\ref{f5}. The values of $U_{\rm ph}$ are given  in units of $\chi
U_{c}$ (per unit area), where $U_c=\pi e^2 \rho_0^2/2\varepsilon
q$ is the Coulomb energy at $a/l\gg 1$
Our calculations show that at $a/l>0.23$ the global minima are
reached for two directions symmetrically deviated  from the [110]
axis on a small angle. In this case one can expect that a domain
structure is formed, and, in average, the system should
demonstrate the minimum resistance in the [110] direction and the
maximum resistance in the perpendicular direction. The
calculations predict the largest resistance anisotropy at
$a/l=0.4$.  If there is no other sources for the native
anisotropy, the high and low resistance directions may alternate
at $a/l\approx 0.23$. At $a/l>1$ the interference term becomes
exponentially small and the 4-fold symmetry is restored.

For $\rho_0$ of order of the average density of electrons at the
high half-filled Landau level and $a/l=0.4$ we estimate the
absolute value of the anisotropy (the energy difference of the
[110] and [$\bar{1}$10] oriented stripes) as 0.7 mK per electron,
which is comparable with the values given by the other possible
mechanism of the anisotropy \cite{10}. The mechanism \cite{10} is
connected with the effective mass anisotropy caused by the
asymmetry of potential confining the electron in the quantum well.
From our estimate we conclude that both mechanisms may work in
parallel and in case of the symmetric confining potential
\cite{11} the native anisotropy can be accounted for the phonon
mechanism. Note that the effect considered depends significantly
on the distance between the electron layer and the surface.
\begin{figure}
\begin{center}
\centerline{\epsfig{figure=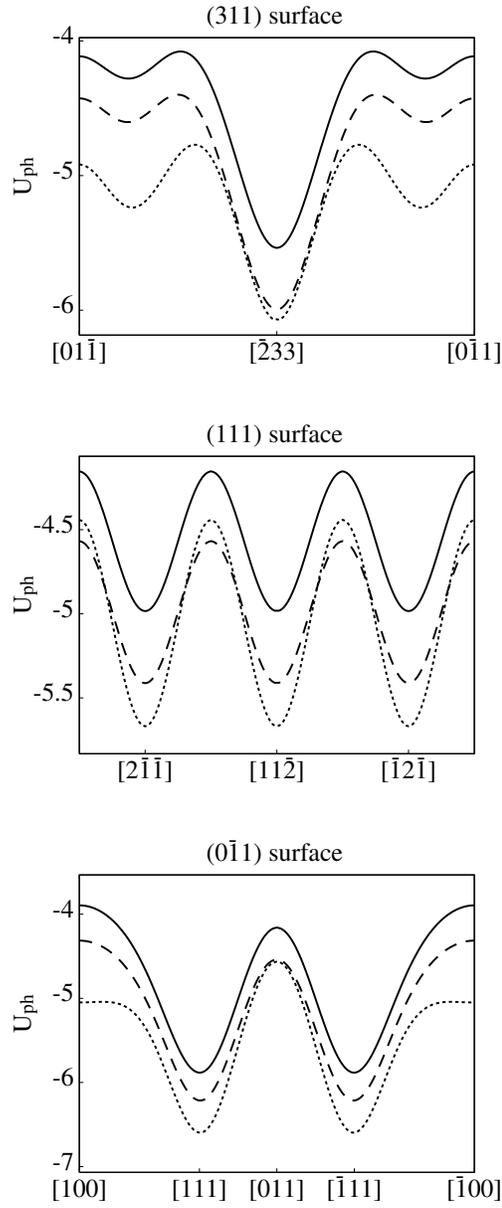,width=8cm}} %\epsfbox{s111.ps}
\end{center}
\caption{\label{f6} Stripe energy anisotropy  for the (111),
(0$\bar{1}$1) and (311) surfaces. Solid curve  - $a/l=1$, dashed
curve - $a/l=0.4$, dotted curve - $a/l=0.2$. $U_{\rm ph}$ is in
$\chi U_c$ units.}
\end{figure}

Assuming the phonon mechanism plays am important role in the
orientational pinning of the stripes,  it is interesting to
analyze the orientation in electron layers parallel to some other
crystallography planes. The results of the calculations for the
(111), (0$\bar{1}$1) and (311) plane are shown in Fig. \ref{f6}.
In these cases, as it follows from the results presented, the
orientation of the stripes does not depend on the ratio $a/l$. For
the (311) surface the phonon mechanism predicts the [$\bar{2}$33]
orientation. This prediction is in agreement with the experiment
\cite{7}. The absolute value of the anisotropy for the (311)
surface is in several times large then for the (001) one. For the
(111) plane the global minima correspond to three different
directions of the stripes ([2$\bar{1}\bar{1}$],
[$\bar{1}2\bar{1}$] and [$\bar{1}\bar{1}2]$). In this case the
polydomain structures may not show any resistance anisotropy. For
the (0$\bar{1}$1) plane the domains with [111] and [$\bar{1}$11]
orientation are energetically preferable, and the polydomain
structure will show the resistance anisotropy with the [011] low
resistance direction. The anisotropy is expected to be small due
to a large angle ($\approx 70^o$) between two preferable
orientations.

\begin{figure}
\begin{center}
\centerline{\epsfig{figure=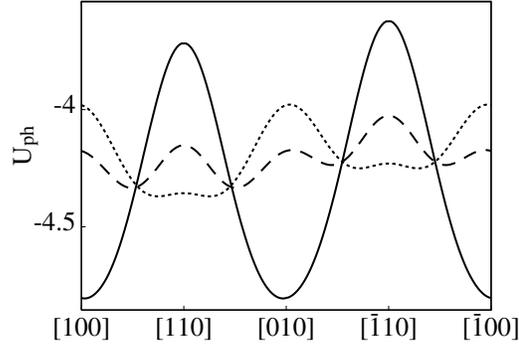,width=8cm}}
%\epsfbox{s001bl.ps}
\end{center}
\caption{\label{f9} Stripe energy anisotropy (in $\chi U_c$ per
one layer) for the bilayer system near the (001) surface at
a/l=0.4. Solid curve - $d/l=0.5$, dashed curve - $d/l=1$, dotted
curve - $d/l=2$}
\end{figure}

To complete our study we consider the orientation of stripes in
the bilayer systems. Let there are two electron layers on the
distances $a$ and $a+d$ from the surface, and the electron density
modulation in the one layer is $\rho_1=\rho_0 \cos {\bf qr}_{\rm
pl}$, and in the other layer is $\rho_2=-\rho_0 \cos {\bf qr}_{\rm
pl}$. For the (001) surface the change of the anisotropy under
variation of the ratio $d/l$ is shown in Fig. \ref{f9}. On can see
that in the bilayer systems with a polydomain structure the angle
between the directions of stripes in different domains becomes
larger for smaller $d/l$. At $d/l=0.5$ these directions are almost
perpendicular to each other. It means that the resistance
anisotropy becomes very small. For the (111), (0$\bar{1}$1) and
(311) crystallography planes we do not find any significant
changes in the stripe orientation under the variation of the
parameter $d/l$. In these cases the bilayer and monolayer systems
will demonstrate the same orientation.

To conclude this section, we would like to point out on the
questions, which, in our opinion, require further experimental
investigation. 1. Is there a dependence of the resistance
anisotropy on the distance between the electron layer and the
surface?  2. Do the electron layers on other interfaces
demonstrate the resistance anisotropy? 3. Is this effect modified
in bilayer systems? Experimental study of these question may be
important for  establishing of the mechanism for the orientational
symmetry breaking in high Landau levels.

\end{document}